\documentclass[aps,prd]{revtex4}

\usepackage{graphicx}

\begin{document}

\title{Ground level muons in coincidence with the solar flare of April 15, 
2001.}

\author{J. Poirier and C. D'Andrea}

\affiliation{Center for Astrophysics, Department of Physics, 
University of Notre Dame, Notre Dame, Indiana 46556 USA.}

\begin{abstract}

The counting rate of single muon tracks from the 
Project GRAND proportional wire chamber array is 
examined during the 
Ground Level Event (GLE) of April 15, 2001.   
The GLE was seen by neutron monitor stations shortly after the time of the 
solar X-ray flare.  
GRAND's single muon data are presented 
and compared with neutron monitor data from Climax, Newark, and Oulu.  
The single muon data 
have mean primary hadron 
energies higher than those of these
neutron monitor stations and so contain information about higher energy 
hadrons.   
For the single muon data for Project GRAND,
the GLE is detected at a statistical significance of 6.1$\sigma$.  
\end{abstract}

\maketitle
\section{Introduction}

On April 15, 2001 the IPS Radio and Space Services website reported an X14 
solar flare with a maximum X-ray intensity at 13.83 UT (8.83 hours EST or 
local time) [{\it IPS Space Services}, 2001].  The flare originated on the 
sun at coordinates S20 W85 
[{\it NOAA Space Environment Center Website}, 2001] and had an 
X-ray intensity of $14 \times 10^{-4}$ W/m$^2$ at one AU 
from the sun.  
Flares of this high intensity are rare. During the period from 
1976 - 1998, only sixteen flares of magnitude X10 or greater were reported.  
Observations on some of these high energy flares are given in [{\it Bieber et al.}, 
2002], [{\it Falcone}, 1999] and [{\it Swinson and Shea}, 1990]. 

Data from the Project GRAND proportional wire chamber array are examined 
for a signal during the time of the 
Ground Level Event (GLE) from this flare.  
During the peak time of the flare, the sun was at an 
elevation angle of $35^\circ$ 
above GRAND's local horizontal plane and an azimuth of $110^\circ$ from north.  
Even though the sun is near the minimum detectable elevation angle for  
GRAND, the Interplanetary Magnetic Field (IMF)   
and the magnetic field of the earth alter 
the direction of the protons from the sun.  
Charged particles from the sun follow a spiral path about the IMF (which itself curves in an Archimedes 
spiral about the sun) and are additionally deflected on entering the 
earth's magnetic field.  
  
Therefore the GLE associated with a particular X-ray flare should occur 
at a slightly later time  
than the X-ray event (due to the longer path length) and at a different 
angle than the sun, thus providing better 
detector acceptance even though a direct line to the sun is at a low 
elevation angle (see also [{\it Munakata et al.}, 2001]). 

GRAND detects ground level muons originating from 
hadronic primaries with energies above the characteristic energy of 
neutron monitors. The value of this characteristic energy depends upon 
the spectral 
index of the proton flux from the sun during this GLE.  The energy response 
function of Project GRAND is discussed in more detail in Section 5. 
Typically, surface muon telescopes are sensitive to proton primary 
energies above about 4 GeV [{\it Shea and Smart}, 2002]. 

\section{Experimental Array}
 
Project GRAND is an array of 64 proportional wire stations located at 
$41.7^\circ$~N and $86.2^\circ$~W at an altitude of 220~m above sea level.  
Each station contains four pairs of 
orthogonal proportional wire chamber planes; each plane of the pair has 80 
wire cells 14~mm wide by 19 mm high. 
Each pair of planes is positioned 200~mm vertically above its neighbor.  
This geometrical arrangement allows for the direction of a muon 
track to be measured to within 0.26$^\circ$, on average, in each of two 
projected angles: $\Theta_{xz}$ and $\Theta_{yz}$.  
The angle $\Theta_{xz}$ is measured in the xz-plane from the zenith (z) toward the east (x), 
$\Theta_{yz}$ is measured in the yz-plane from the zenith toward the north (y).  
A 50~mm thick steel plate is located above the bottom pair of PWC planes.  
The steel plate, in conjunction with the PWC pair of planes  
underneath, enables secondary muon tracks to be distinguished from those of 
electrons or hadrons.   
Each station has a total active area of 1.29~m$^2$; the array currently collects 
data at a rate of $\sim$~2400 muons per second.
 
These data are read from the 64 stations in parallel to a central data 
acquisition system in 70 microseconds.  
Eight computer nodes are used, 
each receiving an event (data from all 64 stations) in sequence.  
The node's task is to search the 80 wires in each of 512 proportional 
wire planes and record the wire numbers in the eight planes of a  
station which had one and only one hit in each plane of that station. 
The eight computer nodes operating 
in sequence are able to conduct 
this search with little overall deadtime.     
Once a particular node has accumulated data on 900 muons in its buffer 
memory, the entire buffer is written to magnetic tape as a single record. 
Upon later off-line analysis, 96\% of this recorded data turn out 
to fit straight tracks.   
A radio signal from WWVB in Boulder, Colorado provides time information 
with millisecond precision.  
This signal is used to determine the start time for the first and the 
last muon event 
in each record.  These times are then recorded along with the muon 
data for that record on magnetic tape.  
Additional details on the Project GRAND detector 
are available on GRAND's webpage [{\it Project GRAND}, 2001].  Other muon 
detectors are described, for example, in [{\it Manakata et al.}, 2001] and [{\it Duldig}, 
2000].

\section{Data Analysis}
 
Project GRAND recorded a continuous data file on magnetic tape containing
information from  April 14, 2001 at 8:40~UT through April 16 at 21:30~UT, 
providing background information before and after the flare time.  
The counting rate from 14.5 hours to 24.5 hours UT was examined in 0.1 
hour bins. The data file is segmented into records, each 
containing information on 900 muons.    
In order to find the time of a particular muon, 
the end time of the current record is compared to the 
end time of the previous record from that computer node; 
the times for each of the 900 muons is interpolated evenly within that interval.   
This interpolation of time should be quite adequate since the average length 
of a record was 3.6 seconds and thus the 
0.1 hour bin size for the data is 90,000 times 
longer than the average time between muons.         

A station can turn on or off, either deliberately for miscellaneous 
repairs (which were not always recorded), or due to problems like, for 
example, 
high humidity coupled with a malfunctioning dehumidifier in a station.     
Deviations (r.m.s.) from average counting rate 
were measured individually for each station of the array.  
In order to prevent spurious individual station variations from 
giving a possible signal in the sum-of-all-stations, 
one-third of the stations with the highest individual r.m.s. deviations 
were eliminated from consideration leaving 39 stations
(an extremely severe cut to ensure with great confidence that 
erratic station behavior could not cause a false signal).
GRAND's counting rate above background is shown as a function of time 
in Figure 1 (lower points with error bars) with the data binned in 0.1 
hour intervals.           

The six bins of time from 14.0 to 14.6 hours UT were 
used as a signal time and 
background bins from 9.5 to 19.5 hours UT (minus the signal time) were used.  
Since it is known that the muon rate depends slightly on the time of day 
[{\it Poirier and D'Andrea}, 2001a] 
(due to small solar effects in air pressure, temperature, and in the IMF), the 
counting rate for the background time period was fit to a curve of the 
form $A(1+Bt+Ct^2)$ 
with A = 496000, B = 0.00101 and C = 0.000241 where t is time measured 
from 14.3 hours UT.  
The smallness of the B and C 
coefficients compared to one is a measure of the flatness of the 
muon counting rate during this time interval.  
The data 
as percentage above background are shown in Figure 1 with error bars.  

For comparison, the data from neutron monitors at Newark for this GLE [{ \it J. Clem 
and R. Pyle}, 2001] and Oulu 
[{\it Oulu Neutron Monitor}, 2001] are also shown.
The Climax neutron monitor data [{\it C. Lopate}, 2001] 
are almost indistinguishable from that of Newark for this 
GLE and, for clarity, are not shown.  

GRAND's muon data rate, although 
sensitive to the atmospheric pressure, has not been corrected 
for changes in pressure. However, 
the pressure was recorded every half hour and found to be 
constant for the entire interval of time analyzed to within $\pm$0.15\%.  
Thus, the pressure is unlikely to be a factor in the analysis.  
Indeed, during the time of the 
flare the pressure actually rose slightly which would, if anything, depress 
the counting rate slightly during the time of the GLE.  Thus, pressure variations 
could not be a cause of the excess counting rate observed from 14.0 to 14.6 
hours UT.

The muon counting rate for April 14 and April 16 (the preceding 
and the next day from the signal) 
are shown in Figure 2 as well as the 
data containing the signal on April 15.  
The background curve fit 
to the April 15 data is also included.    
It is seen that the background from the day before and the 
day after are similar to the 
background curve for April 15; 
also, the average of four years of data show a similar shape.  
The zero of muon counting rate is suppressed in order to better show 
these small deviations from an almost constant rate. 
%
%============fig. 1

\begin{figure*}[h]
\includegraphics*[width=11.0cm]{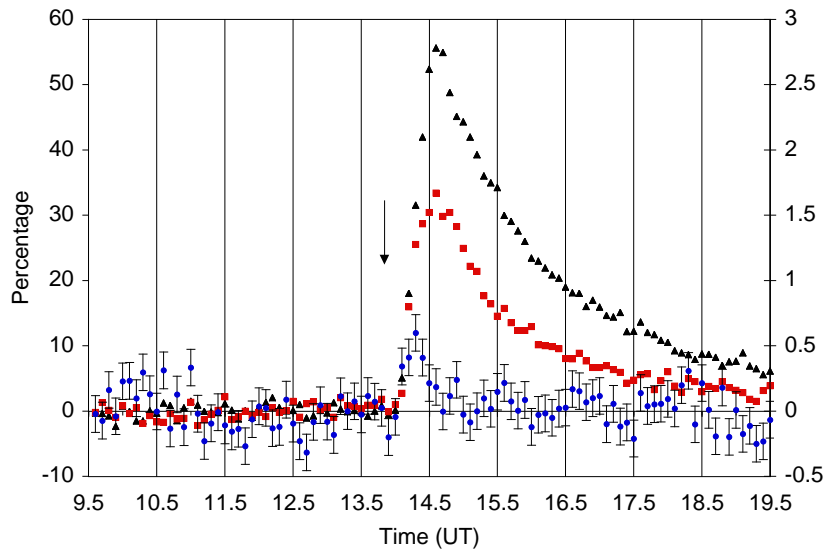}
\caption{Percentage counting rate increase above background versus time 
on April 15.  
Bottom points (circles) with error bars are from GRAND
with \%-scale on the right; background is the curve in Fig. 2.
Middle points (squares) are from Newark Neutron Monitor (left scale). Top points 
(triangles) are from Oulu Neutron Monitor (left scale).  
The arrow denotes the onset of the X-ray signal on the sun (at 13.83 hours UT 
[{\it IPS Space Services}, 2001]). }  
\end{figure*}

%

%=========fig 2
\begin{figure*}[h]
\includegraphics*[width=11.0cm]{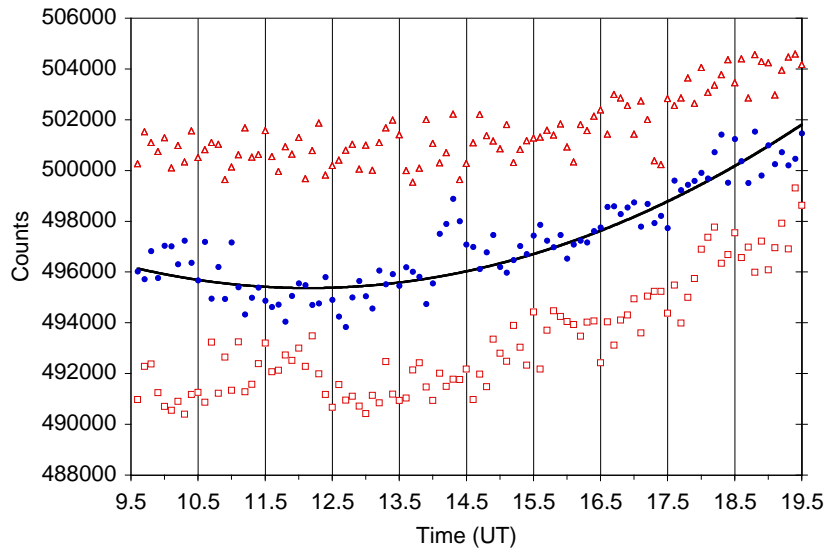}
\caption{Single muon counting rates in 0.1 hour bins 
versus time with suppressed zero to emphasize the small anisotropies.
Bottom points (squares) are April 14.  
Middle points (circles) are April 15 with background 
curve fit and correspond to the numbers on the left.  
Top points (triangles) are April 16.  The April 14 and 16 data are offset 
for clarity.  
Note that the small hourly background variations in time ($\pm0.6\%$)
are consistent over this 
three-day period.}
\end{figure*}

The amount of signal above background was determined by the excess 
of the six signal bins above the fitted background curve.
The signal in these six bins yields 10,708 counts above 
background with an error of 1762; this error includes the error on 
the background fit (A) and uses the r.m.s. deviation of the background 
about the fitted curve as an estimator for the rms error of the 
data in the signal region.  This 
yields a Ground Level Event signal with a significance of 6.1$\sigma$ 
(where $\sigma$ is the error on the signal).
 
Each of the 39 stations has an active area of 1.29 m$^2$; 
correcting for those stations which were not 100\% efficient 
yields a total effective area of 42 m$^2$ for this analysis.  
The excess counting rate was 255 muons/m$^2$ or 0.36\% of the background 
muon counting rate during the 0.6-hour signal interval.

\subsection{Can GRAND see gamma ray or neutron primaries from this GLE?}
To determine if GRAND detects secondary muons from gamma ray [{\it Poirier et al.}, 2001] 
or neutron primaries which originate from the sun (rather than the protons studied above),
the muon rate was analyzed inside a $12^\circ$ by $12^\circ$ square angular window
centered on the sun.  
During the signal time from 14.0 to 14.6 hours UT, 
368 counts were measured compared to 
an average of 376 background counts for the 0.6 hour interval of time 
as obtained during 9.5 to 19.5 hours UT (minus the signal time).    
The low counting rate in this angular window is caused by the sun's low 
elevation angle at that time.   
A negative signal of --8 counts from background is observed inside this $12^\circ$ by 
$12^\circ$ window, or --0.4~$\sigma$ if 
compared to the error on the signal.  
Thus there is no evidence for direct gamma rays or neutrons from the sun; as well, they 
cannot be the cause of the 6.1~$\sigma$ excess obtained in the data with 
no angular cuts. 

\section{Angular Distribution}  

The sky was divided into sixteen angular regions each with approximately 
equal counting rates in order to obtain some information about 
the angular characteristics of the signal [{\it Poirier and D'Andrea}, 
2001b].  
Each individual region of the sky was examined for the significance of 
a possible GLE signal. 
The significance of the GLE signal (in terms of sigma) 
for each of the 16 windows is shown in Table 1.   
For these 16 calculations, the background was individually 
fit to the same quadratic form as before with coefficients A, B, and 
C; however, due to the reduced amount of data in a region ($\sim$1/16th 
of the entire angular region), 
the coefficients B and C were fixed at the value obtained for the 
whole sky and only the A parameter was varied in the fit to estimate the 
background under the signal region for each of the 16 angular regions.  The angular information presented in Table 1 could be used in later 
analyses in an overall fit of this GLE using data from other muon detectors 
and neutron monitor stations around the world to obtain the mean arrival 
direction, pitch angle distribution, and spectral index.  

\begin{table}[bhtp]
\caption{
The significance (in terms of sigma) of the GLE signal 
for each region of the sky from  
14.0 to 14.6 UT above the background measured from 9.5 to 19.5 UT.   
$\Theta_{xz}$ is measured from the zenith toward the east; 
$\Theta_{yz}$ is measured from 
the zenith toward the north. 
Local zenith is at the center of the table; to the right is east, 
to the top is north.}

\begin{center}
\begin{tabular} {|c||c|c|c|c|} \hline \hline
$\Theta_{yz}$
 & \multicolumn{4}{|c|} {$\Theta_{xz}$}  \\
\hline \hline
 & $\le$ --13$^\circ$ & --13 to 0 & 
  0 to 13 & $\ge$13$^\circ$ \\
\hline
$\ge$13$^\circ$ & 1.1 & 0.4 & 0.6 & 0.5 \\
\hline
0 to 13 & 1.4 & 2.5 & 3.7 & 1.7 \\
\hline
--13 to 0 & 1.4 & 2.3 & 3.3 & 0.9 \\
\hline
$\le$ --13$^\circ$ & --0.8 & --0.1 & --0.1 & --1.0 \\
\hline \hline
\end {tabular}
\end {center}
\end {table}

An overall fit for each hemisphere (in a local horizon coordinate system) 
gives a signal strength of 4.8~$\sigma$ for the 
northern and 1.9~$\sigma$ for the southern hemisphere, while a signal 
strength of 3.7~$\sigma$ is obtained for the eastern and 3.4~$\sigma$ 
for the western hemisphere.  These results show a definite preference for the 
northern hemisphere compared to the southern and only a slight preference 
for the eastern hemishpere over the western.   
The mean location of the sun at GRAND during the time of the GLE was 
$\theta_{xz}$ = +53$^\circ$ (eastward) 
and $\theta_{yz}$ = --26$^\circ$ (southward). 

\section{Primary Energy Sensitivity}

The response function of GRAND as a function of energy for primary protons 
yielding muons which reach detection level is calculated by using
the Monte Carlo program FLUKA [{\it Fass\'o et al.}, 2000].  
The atmosphere in these simulations is approximated by
50 layers of air beginning 80~km above sea level.  The calculation is 
similar to that reported
in [{\it Poirier et al.}, 2001] except that the primary particles were 
changed from the gamma
rays of that reference to primary protons for this calculation.
Each primary proton is 
incident on the top of the atmosphere at zero degrees from zenith angle 
(as well as three non-zero angles for the 3, 10, and 30 GeV energies) 
and all non-hadronic secondaries are followed until they are below 
the threshold 
energy to produce pions or the charged tracks reach ground level.  
The calculation is done for
a series of fixed proton primary energies for energies from 1~GeV to 1~TeV.
A summary of the results of these calculations is contained in
Table 2 for proton primaries incident at the top of the atmosphere at 
0$^\circ$ from zenith.
Figure 3 is a plot of this response function.  Protons incident at 
0$^\circ$ from zenith are shown by a dashed line.  Incident angles of 
17$^\circ$, 34$^\circ$, and 51$^\circ$ 
from zenith (solid lines) show GRAND has progressively smaller responses 
to angles 
which deviate from normal incidence.  
The deviation is fractionally larger at the lowest energy (3 GeV) plotted.

 L. Miroshnichenko [{\it Miroshnichenko}, 2002 {\it 
preliminary}] has calculated a preliminary fit to monitor stations 
at Apatity, Moscow (IZMIRAN), Rome, and Athens to obtain the integral flux of 
protons from this GLE versus the energy of the protons from 1 to 10 GeV.  
These values are plotted in Figure 4 as square points.
The circular points below 0.1 GeV are proton fluxes obtained by the GOES 
satellite [{\it GOES Website}, 2001]..  

To obtain the overall differential response expected of GRAND for this GLE, 
the response function for GRAND (at 0$^\circ$ in Table 2) is plotted 
in Figure 4 as a dashed 
line (the numbers on the ordinate are the same with units of muons at 
detection level per proton of that primary energy). 
This response function is then folded with the flux of primary protons in 
each energy interval to obtain the differential energy response of GRAND 
for this GLE (shown as diamond points  below 10~GeV).  
At energies  $\geq10$~GeV, the proton intensity (shown by solid 
curves) was 
approximated using a power law spectrum 
with differential spectral indices of --8 (upper curve) and --10 (lower curve) which are normalized 
to the point at 9~GeV.  

\begin{figure}
\includegraphics[width=8.3cm]{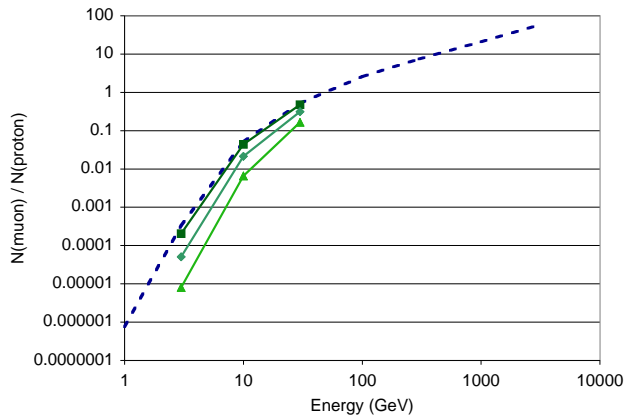}
\caption{GRAND's response as a function of energy.  The dashed line is the 
number of muons at detection level per incident 
proton primary.  The highest solid line is for protons 
incident at 17$^\circ$ from zenith, the middle line is for protons at 34$^\circ$,
and the bottom line is protons at 51$^\circ$.
}
\end{figure}.  

\begin {table}
\caption{The results of the Monte Carlo calculation with the FLUKA code for
the number of primary protons ($N_p$) of kinetic energy $E_p$ 
incident normally at the top of the atmosphere yielding the number 
of muons ($N_{\mu}$) which reach 
ground level for different discrete primary proton kinetic energies.}
\begin{center}
\begin{tabular} {|c|c|c|c|} \hline \hline
$E_p$ (GeV) & $N_p$ & $N_{\mu}$ & $R=N_{\mu}$/$N_p$ \\
\hline
1 & $36.0\times10^6$ & 27 & $7.5 \times 10^{-7}$ \\
\hline
3 & $2.00\times10^6$ & 664 & $3.32 \times 10^{-4}$ \\
\hline
10 & $0.75\times10^6$ & 38511 & 0.0513 \\
\hline
30 & $0.40\times10^6$ & 208928 & 0.522 \\
\hline
100 & $0.10\times10^6$ & 256847 & 2.57 \\
\hline
300 & $0.10\times10^6$ & 738561 & 7.39 \\
\hline
1000 & $0.040\times10^6$ & 825910 & 20.6 \\
\hline \hline
\end {tabular}
\end {center}
\end {table}

\begin{figure*} 
\includegraphics[width=11.0cm]{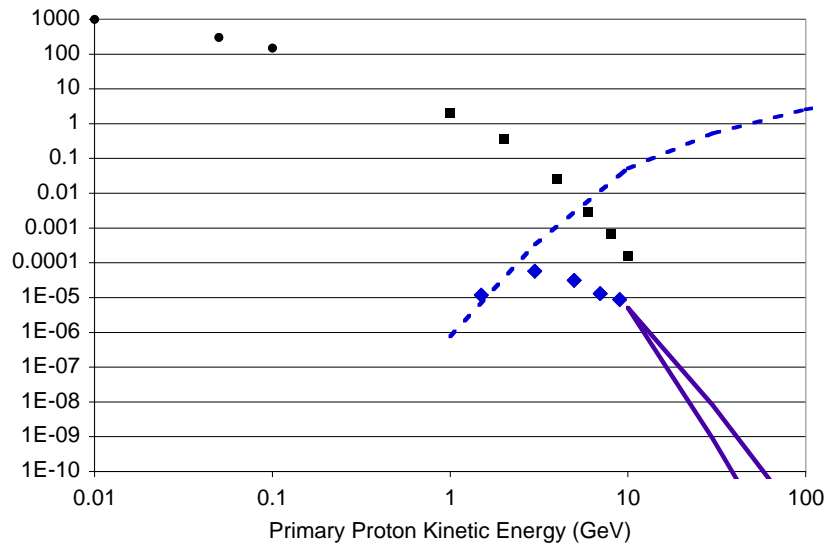}
\caption{
The circular points represent integral proton intensity data 
from the GOES satellite 
[{\it GOES Website}, 2001] in units of 
particles $cm^{-2} s^{-1} sr^{-1}$, 
while the square points are integral proton intensities 
from neutron monitor data compiled by 
Leonty Miroshnichenko [{\it Miroshnichenko}, 2002 {\it preliminary}] in 
the same units with an estimated factor of two error.  
The dashed line shows the R-values 
from Table 2 versus the primary proton 
kinetic energy calculated with FLUKA (in units of muons per primary 
proton).    
The diamond points from 1.5 to 9~GeV represent the fold of 
proton intensity with R-values to obtain the expected muon 
differential intensity at ground level for each interval (in units of 
particles $cm^{-2} s^{-1} sr^{-1} GeV^{-1}$. 
At energies greater than 10~GeV, the spectrum has been 
approximated with differential spectral indices of --8 (upper solid curve) and 
--10 (lower solid curve) normalized to the 9~GeV point.
}
\end{figure*}

\subsection{Asymptotic Direction of Observation versus Rigidity}
For rigidities (momentum per charge) in the GV range (kinetic energies in 
the GeV range), the 
primaries suffer  
deflections in the earth's magnetic field.  
In order to place GRAND's data for this GLE in the perspective of other 
muon stations and neutron monitors, it is of interest to know the 
direction of a proton primary before entering 
the earth's magnetic field (asymptotic direction) 
in order to arrive at vertical incidence 20~km above Project GRAND.  
This asymptotic direction depends on the rigidity of the primary as well as the 
location of the detector on the surface of the earth.  
We are indebted to Margaret Shea and Don Smart for performing this 
calculation [{\it Shea and Smart}, 2002], 
summarized in Figure 5, which shows the asymptotic 
direction of an antiproton starting at 20 km above sea level at the 
geographical location of GRAND (41.7$^\circ$~N and 86.2$^\circ$~W) and 
integrating the trajectories of various rigidities outward through a model 
of the earth's 
magnetic field to find their direction after exiting this field.  
The coordinate system for the asymptotic directions can be 
visualized as the intersection of this antiproton's trajectory 
on a huge sphere of the earth's coordinates with a radius much 
larger than the earth's magnetosphere.     
This is then the direction of a {\it proton} traveling 
{\it toward} the earth in the geomagnetic field which 
would reach an altitude of 20~km height above GRAND at 
0$^\circ$ relative to its local zenith.  

The model of the earth's magnetic field which was used is the International 
Geomagnetic Reference Field, Epoch 1995 (I95). 
This is a quiescent field model and does not take into
account magnetospheric perturbations (the $K_p$ index during the GLE was 4).
The individual asymptotic directions change with time because of the
magnetospheric configuration.  However, for rigidity values above about
2 to 2.5 GV, the changes are slight.  Of greater importance would be
changes related to the secular variation of the geomagnetic field which, for
some areas of the world, can greatly alter the asymptotic directions close
to cutoff values.  Also during major geomagnetic disturbances the asymptotic
directions can be changed considerably since the cutoff rigidity can be
considerably lowered.  However, for the Easter Sunday GLE and for GRAND,
the values calculated should be adequate for almost any analysis of muon 
response [{\it Shea and Smart}, 2002].

\begin{figure*}[h]
\includegraphics*[width=11.0cm]{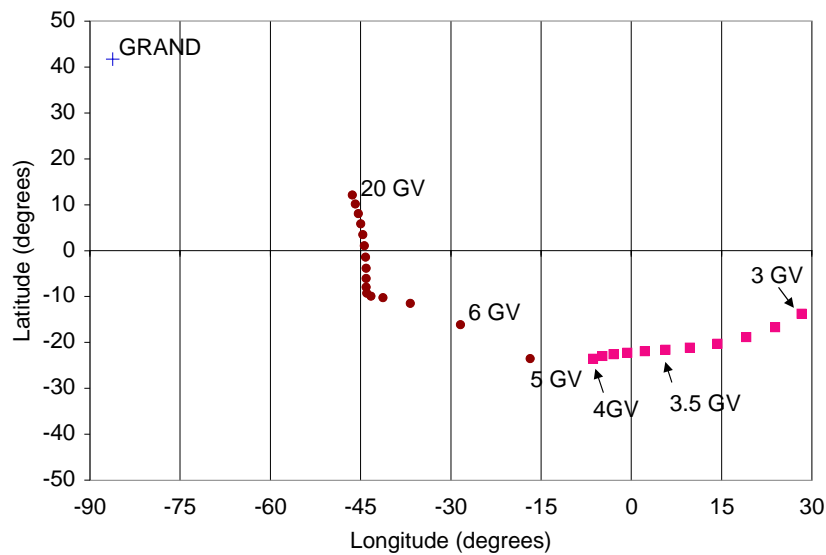}
\caption{Asymptotic directions as a function of particle rigidities for GRAND
[{\it Shea and Smart}, 2002].  
}
\end{figure*}

Data on the Interplanetary Magnetic Field (IMF) is shown in figure 6 [{\it GOES 
website}, 2002].  The magnitude of the IMF is shown as circular points.  
The components in GSE coordinates are by squares (x, from the earth to the 
Sun along the ecliptic), triangles (y, perpendicular to x in the ecliptic 
plane), and diamonds (z, in the direction of the north ecliptic pole).  
While the magnitude of the IMF remains fairly constant, the direction of 
the magnetic field is changing over the period of this GLE.

\begin{figure*}[h]
\includegraphics*[width=11.0cm]{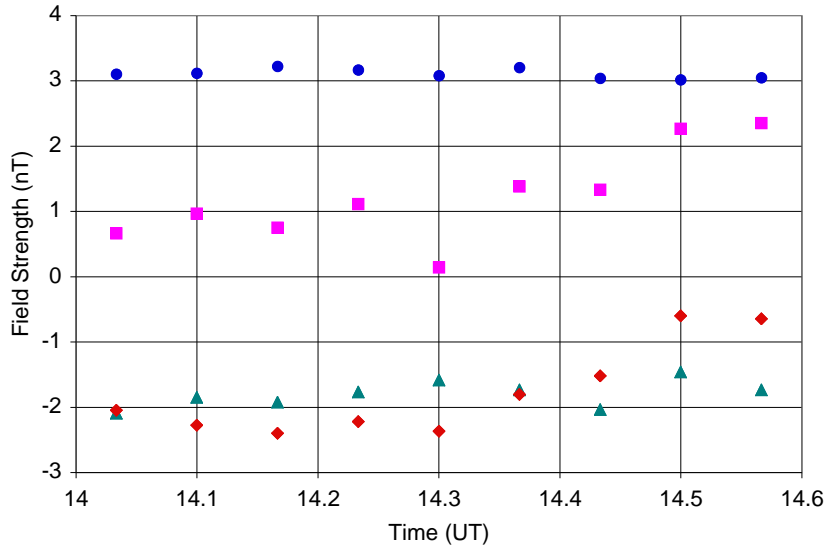}
\caption{
Interplanetary Magnetic Field during the April 15, 2001 GLE [{\it GOES 
Website}, 2002].  The circular points represent the magnitude of the 
magnetic field.  The squares, triangles, and diamonds represent the x,y, 
and z components of the magnetic field in the GSE coordinate system.
}
\end{figure*}

\section{Other Observations}

Data from the Climax, Newark, and Oulu Neutron Monitor Stations 
are compared for this time interval on April 15 [{\it Lopate, Clem and Pyle, and Oulu 
Neutron Monitor Website}, 2001].  In Figure 1, the data 
for Climax are very close to that of Newark and for clarity are not shown.
The energy thresholds for the 
Climax, Newark, and Oulu neutron monitor stations are influenced by their 
vertical geomagnetic cutoff 
rigidity at 3.0, 2.1, and 0.8 GV respectively [{\it University of Chicago 
Neutron Monitor Website}, 2001] and overburden of air; 
their corresponding heights above sea level are 3400, 50, and 15~m.  
The corresponding numbers for GRAND are a geomagnetic cutoff rigidity 
of 1.9~GV and a 
height of 220~m above sea level.  However, the primary energy for 
GRAND is influenced by the different mechanisms which produce muons 
(rather than neutrons) at detection level. 
While looking at single muon tracks at ground level, 
GRAND's primary proton energy 
is approximately 4~GeV (see Figure 4 and [{\it Shea and Smart}, 2002]) for 
the spectral shape of this GLE.
     
When binned in six-minute intervals, the 
Newark signal has an onset time at 14.21 hours, the Climax signal  
at 14.16 hours, and Oulu at 14.27 hours 
(where the ``onset" is the time of the half-height of the 
signal as measured from its rise above background to 
the highest measured data point of the GLE).    
A similar calculation for the onset time for GRAND 
is 14.09 hours.  
The width (full width at half-of-maximum above background) 
of the Climax, Newark, and Oulu peaks are 
1.27, 1.24, and 1.61 hours in contrast to GRAND's value of 0.48 hour.  
Thus the GRAND peak is: a) shorter in duration (typically, higher 
energies have a shorter duration than lower energies as is seen in 
satellite data for protons at lower energies [{\it GOES 
Website}, 2002]), b)
the onset is $\sim$0.1 hour earlier 
(the earlier time might be expected on the basis of 
higher primary energies compared to the neutron monitors [{\it Clem and 
Dorman}, 2000] 
with faster protons traveling 
at smaller pitch angles yielding shorter 
path lengths, and c) smaller in amplitude due to 
GRAND's sensitivity to higher primary energies.

\section{Conclusions}

Project GRAND sees a ground level signal with a significance of 
6.1~$\sigma$ when examining 
the secondary muon counting rate at ground level between 14.0 and 14.6 
hours UT on April 15, 2001.  This signal is obtained 
with no restrictions on the angle of the muons.  
This is consistent with protons 
originating at the surface of the sun and accelerated during the time 
of the X14 X-ray flare.    

A slightly negative signal (--0.4 $\sigma$) 
is obtained if a small region of the sky ($\pm5^\circ$ in $\theta_{xz}$ 
and in $\theta_{yz}$) near the angle of the sun's position is examined.   
Thus, neither gamma ray nor neutron 
primaries are the cause of the 
6.1$\sigma$ signal found in the data with no angular cuts. 

GRAND's GLE detection occurs slightly earlier than Climax, Newark, or Oulu  
neutron monitor signals which might be expected from the 
slightly higher primary energies detected by GRAND (above about 4~GeV) 
compared to the neutron 
monitors (the vertical geomagnetic cutoff energies for the  
Climax, Newark, and Oulu neutron monitor stations are 
3, 2, and 0.8 GeV respectively). Earlier times would be consistent with 
smaller pitch angles relative 
to the IMF (and thus a shorter flight path)  for the higher energies as well
as the increased velocity. The fact that GRAND's GLE 
signal is not as pronounced as those from the neutron monitors 
indicates that there are fewer particles 
at GRAND's higher energies.  The mean energy of primary protons which 
produces the detected muons depends somewhat upon the spectral index of this GLE 
in the 4~GeV region.  In the future, combined analyses of world-wide neutron monitor stations 
and muon telescopes are 
anticipated to yield more detailed information on this GLE.

\begin{acknowledgements}

The authors wish to thank 
Monica Dunford, Jon Vermedahl, and Mari\'e L\'opez del Puerto
for their participation in the data analyses; Kent Doggett 
(and the ACE science team); the GOES science team; and   
Stefan Roesler and Alberto Fasso for their assistance with the Monte 
Carlo simulations.  Thanks to Leonty Miroshnichenko for preliminary 
intensity calculations.   We also thank 
Margaret Shea and Don Smart for calculations of GRAND's asymptotic 
direction as a function of rigidity; 
John Clem and Roger Pyle (and Bartol), Cliff Lopate (and Climax), and the Oulu
Neutron Monitor for the use of their data; Ted Bowen for suggestions on 
improving the text; 
and Jule Poirier for proofreading.
The Newark neutron monitor is operated by the Bartol 
Research Institute Neutron Monitor Program and is funded by National 
Science Foundation Grant ATM-0000315.  The Climax Neutron Monitor is 
operated by The University of Chicago and is funded through National 
Science Foundation Grant ATM-9912341.
Project GRAND is funded through the University of Notre Dame and 
private grants.  
\end{acknowledgements}

\end{document}